\documentclass[a4paper]{jpconf}
\usepackage{cite}
\usepackage{graphicx}

\newcommand{\Ref}[1]{Ref.~\citen{#1}}
\newcommand{\Fig}[1]{Fig.~\ref{#1}}

\newcommand{\EA}[5]{\ensuremath{
\mbox{#1}^{+\mbox{\scriptsize{#2}}}_{-\mbox{\scriptsize{#3}}}
\times\mbox{10}^{#4\mbox{\scriptsize{#5}}}}}

\def\ead#1{\vspace*{5pt}\address{$^*$E-mail (speaker): \mailto{#1}}}

\begin{document}
\title{The large-angle photon veto system for the {NA62} experiment at the {CERN} {SPS}}
\author{F~Ambrosino$^1$, B~Angelucci$^2$, A~Antonelli$^3$, F~Costantini$^2$, 
G~D'Agostini$^4$, D~Di~Filippo$^1$, R~Fantechi$^2$, S~Gallorini$^2$, 
S~Giudici$^2$, E~Leonardi$^4$, I~Mannelli$^2$, P~Massarotti$^1$, 
M~Moulson$^{3,*}$,
M~Napolitano$^1$, V~Palladino$^4$, F~Rafaelli$^2$, M~Raggi$^3$, G~Saracino$^1$,
M~Serra$^4$, T~Spadaro$^3$, P~Valente$^4$, S~Venditti$^2$}

\address{$^1$Dipartimento di Scienze Fisiche dell'Universit\`a and Sezione INFN, Napoli, Italy}
\address{$^2$Dipartimento di Fisica dell'Universit\`a and Sezione INFN, Napoli, Italy}
\address{$^3$Laboratori Nazionali di Frascati dell'INFN, Frascati, Italy}
\address{$^4$Dipartimento di Scienze Fisiche dell'Universit\`a and Sezione INFN, Napoli, Italy}

\ead{Matthew.Moulson@lnf.infn.it}

\begin{abstract}
The branching ratio (BR) for the decay $K^+\to\pi^+\nu\bar{\nu}$ is a 
sensitive probe for new physics. The NA62 experiment at the CERN SPS will 
measure this BR to within about 10\%. To reject the background 
from dominant kaon decays with final state photons, the large-angle photon 
vetoes (LAVs) must detect photons of energy as low as 200 MeV with an 
inefficiency of less than 10$^{-4}$. The LAV detectors make use of lead glass 
blocks recycled from the OPAL electromagnetic calorimeter barrel. We describe 
the mechanical design and challenges faced during construction, the 
characterization of the lead glass blocks and solutions adopted for 
monitoring their performance, and the development of front-end electronics 
to allow simultaneous time and energy measurements over an extended dynamic 
range using the time-over-threshold technique. Our results are based on 
test-beam data and are reproduced by a detailed Monte Carlo simulation 
that includes the readout chain.
\end{abstract}

\section{The NA62 Experiment}

\begin{figure}
\centering
\includegraphics[width=0.35\textheight]{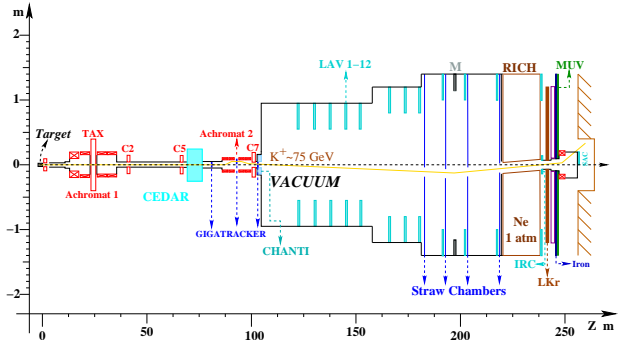}
\caption{The NA62 experimental layout.}
\label{fig:expt}
\end{figure}

The decays $K^+\to\pi^+\nu\bar{\nu}$ and $K_L\to\pi^0\nu\bar{\nu}$ are
flavor-changing neutral-current processes for which the rates are highly 
suppressed in the Standard Model (SM). At the same time, largely 
because the hadronic matrix elements can be obtained from experimental data 
on $K_{\ell3}$ decays, the SM branching ratios 
(BRs) for these decays can be predicted with minimal 
intrinsic uncertainty (see \Ref{C+11:kaonRev} for a recent review).
The $K\to\pi\nu\bar{\nu}$ decays are therefore a sensitive probe of the SM 
flavor sector and provide constraints on the CKM unitarity triangle that are 
complementary to those from measurements of $B$-meson decays. 
On the other hand, the tiny BRs for these decays are notoriously difficult
to measure, not least because of the three-body kinematics with two
undetectable neutrinos in the final state. At present, the experimental value 
of the BR for the decay 
$K^+\to\pi^+\nu\bar{\nu}$ is \EA{1.73}{1.15}{1.05}{-}{10} on the basis 
of seven detected candidate events \cite{E949+08:Kpnn2}.
The goal of NA62, an experiment at the CERN SPS,
is to detect $\sim$100 $K^+\to\pi^+\nu\bar{\nu}$ decays
with a S/B ratio of 10:1 
in two years of data taking beginning in 2014.
The experiment is fully described in \Ref{NA62+10:TDD}.
The experimental layout is illustrated in \Fig{fig:expt}.

NA62 will make use of a 75-GeV unseparated positive secondary beam
with a total rate of nearly 800~MHz, of which $\sim$50 MHz is $K^+$'s.
All 800~MHz of beam particles are tracked by three silicon-pixel 
tracking detectors just upstream 
of the vacuum decay volume, providing measurement of the 
$K^+$ trajectory and momentum.
The vacuum decay volume ($10^{-6}$ mbar) begins $\sim$100~m downstream 
of the production target.   
5~MHz of kaon decays are observed in the 65-m long vacuum decay region.
Large-angle photon vetoes (LAVs) are placed at 11 stations along 
the decay region and provide full coverage for decay photons with
$\mbox{8.5~mrad} < \theta < \mbox{50~mrad}$.
The last 35~m of the decay region hosts a dipole 
spectrometer with four straw-tracker stations operated in vacuum. 
The NA48 liquid-krypton 
calorimeter \cite{NA48+07:NIM} is used to veto high-energy photons at
small angle. 
Additional detectors further downstream extend the coverage of
the photon veto systems (e.g. for particles traveling 
in the beam pipe).

The experiment must be able to reject background from dominant decays 
such as $K^+\to\pi^+\pi^0$ at the level of $10^{12}$. Cuts on the $K^+$ 
and $\pi^+$ momenta provide a rejection factor of $10^4$ and ensure 
that the photons from the $\pi^0$ have 40~GeV of energy. Forward photons
that are intercepted by the LKr calorimeter and small-angle vetoes have
much higher energies than the photons intercepted by the LAVs.
Nevertheless, photons in the LAVs from decays such as 
$K^+ \to \pi^+\pi^0$ may have energies from a few tens of MeV to several
GeV. In order to detect the $\pi^0$ with an inefficiency of $\leq 10^{-8}$,
the maximum tolerable inefficiency in the LAV detectors for photons 
with energies as low as 200~MeV is $10^{-4}$. In addition, the large-angle
vetoes must have good energy and time resolution (10\% and 1~ns for 1 GeV 
photons) and must be compatible with operation in vacuum.

\section{The Large-Angle Veto System}

The NA62 LAV detectors make creative
reuse of lead glass blocks recycled from the OPAL electromagnetic calorimeter
barrel\cite{OPAL+91:NIM}, which became available in 2007 when various 
technologies were under consideration for the construction of the LAV 
detectors. Other solutions considered included a lead/scintillating tile design 
originally proposed for use in the (later canceled) CKM experiment at Fermilab, 
and a lead/scintillating-fiber design based on the electromagnetic calorimeter
for the KLOE experiment. Prototype instruments based on each 
of the three technologies were obtained or constructed, and tested
with the electron beam at the Frascati Beam-Test Facility. 
These tests demonstrated that all three technologies are suitable 
for use in NA62 \cite{A+07:Veto}. In particular, the inefficiency for the 
detection of single, tagged electrons with the OPAL lead glass modules
was measured to be \EA{1.2}{0.9}{0.8}{-}{4} at 203 MeV and
\EA{1.1}{1.9}{0.7}{-}{5} at 483 MeV.
Basing the construction of the LAV system on the OPAL lead glass modules 
provides significant economic advantages.

\begin{figure}
\centering
\includegraphics[height=0.17\textheight]{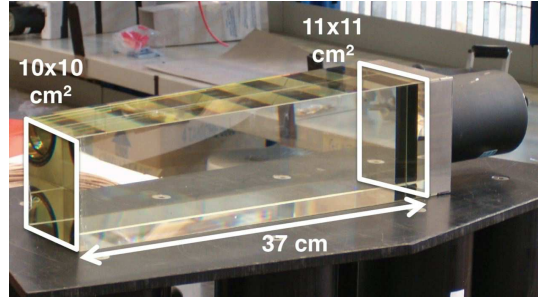}
\caption{A module from the OPAL calorimeter, without wrapping and with 
reinforcement plates at the interface between the glass and the steel 
flange.}
\label{fig:block}
\end{figure}
The modules from the central part of the OPAL 
electromagnetic calorimeter barrel consist of blocks of Schott SF57 lead glass.
This material is about 75\% lead oxide by weight and has a density
$\rho = 5.5~{\rm g}/{\rm cm}^3$ and a radiation length $X_0 = 1.50$~cm;
its index of refraction is $n \approx 1.85$ at $\lambda = 550~{\rm nm}$
and $n \approx 1.91$ at $\lambda = 400~{\rm nm}$.
The Cerenkov light produced by electromagnetic shower particles in the 
lead glass is read out at the back side of the block by a Hamamatsu R2238 
76-mm PMT coupled via a 4-cm long cylindrical light 
guide of SF57 of the same diameter as the PMT.
The rear face of the glass block is glued to a 1-cm thick stainless 
steel flange. 
A mu-metal shield surrounding the PMT and light guide is also glued to the
flange. Figure~\ref{fig:block} shows a picture of a complete module.

\begin{figure}
\centering
\includegraphics[height=0.22\textheight]{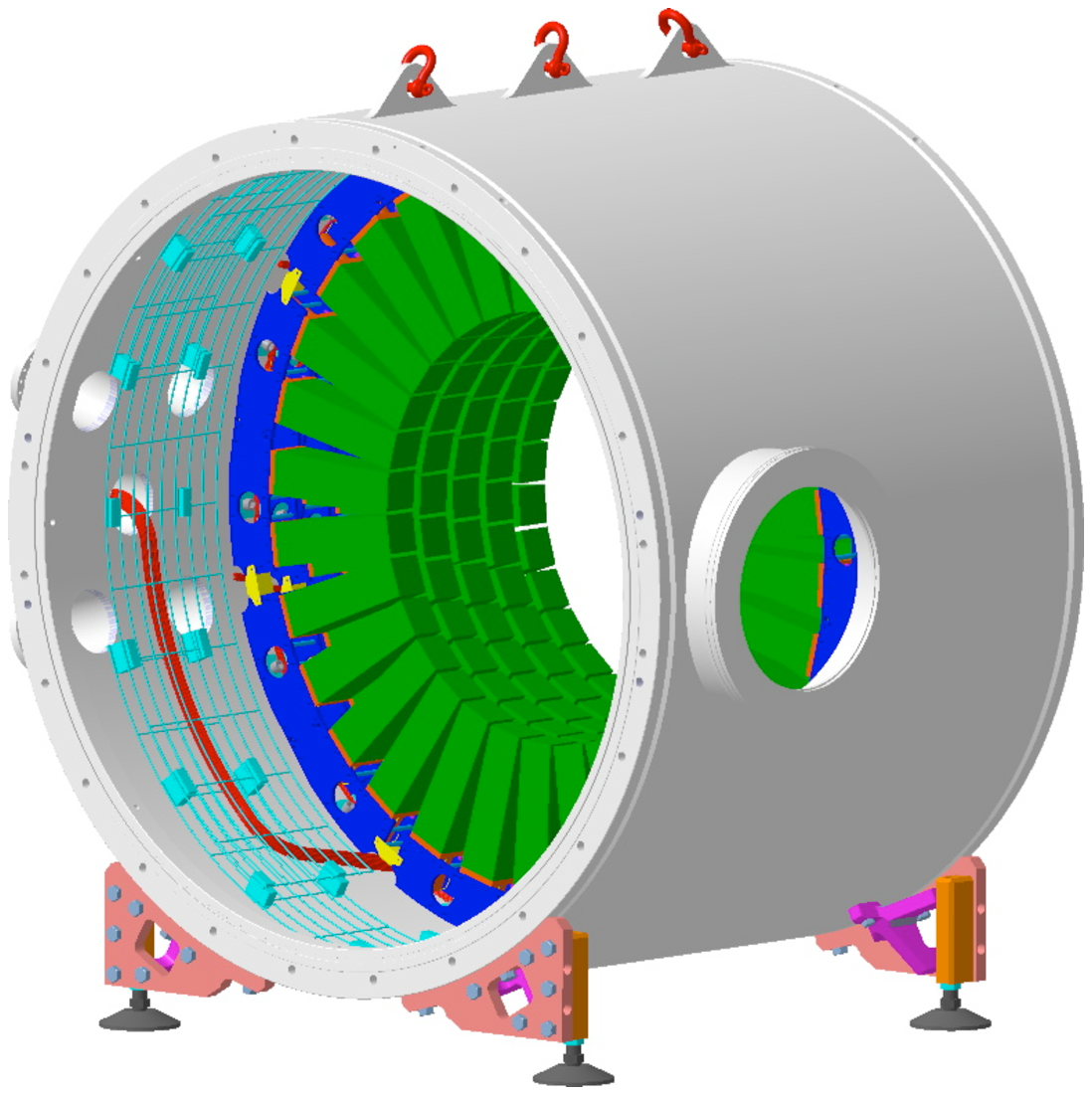}
\hspace{0.10\linewidth}
\includegraphics[height=0.22\textheight]{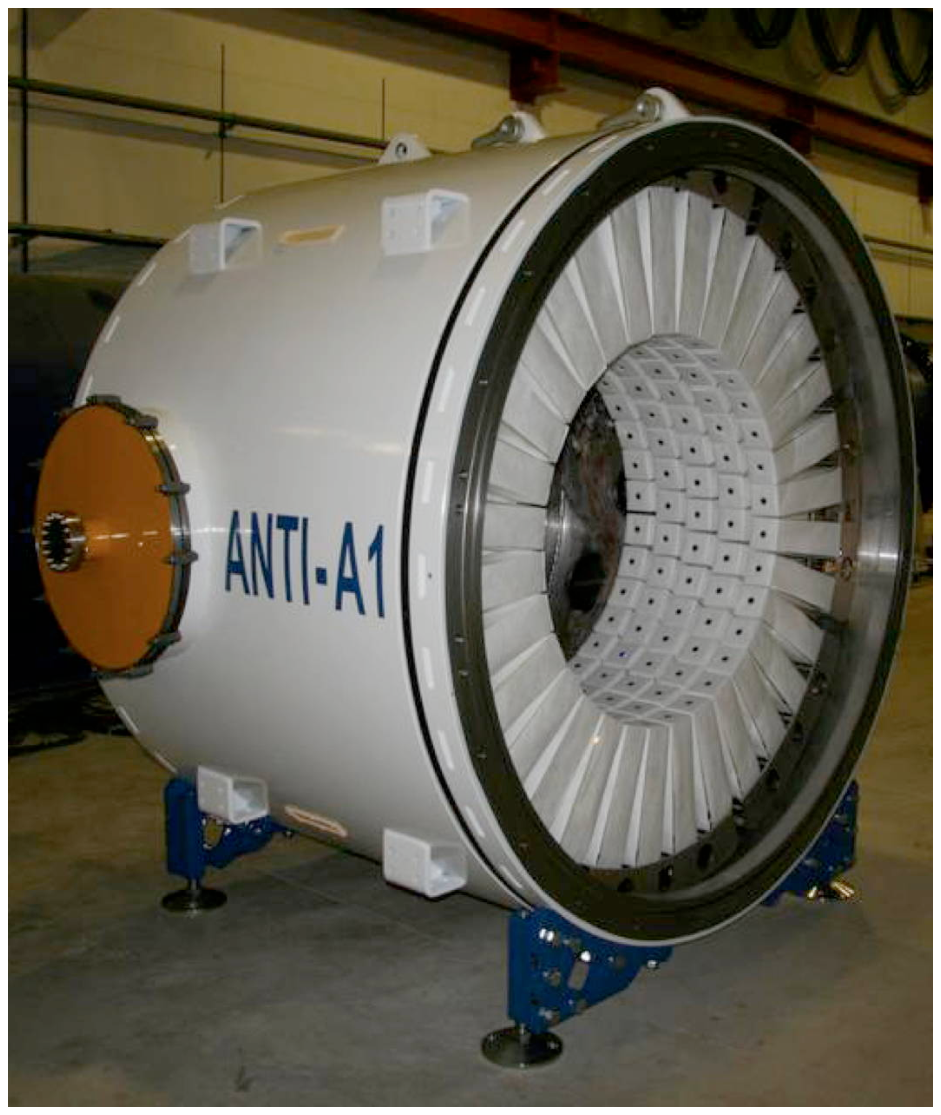}
\caption{Design study (left) and completed prototype A1 veto station (right).}
\label{fig:station}
\end{figure}
A LAV station is made by arranging these blocks around the inside of 
a segment of vacuum tank, with the blocks aligned radially to form an 
inward-facing ring. Multiple rings are used in each station.
The blocks in successive rings are staggered in azimuth; the rings are 
spaced longitudinally by about 1 cm. The LAV system consists of a total of
12 stations.
Stations A1--A5 are 217~cm in diameter and contain 160 blocks arranged in 5 
layers, stations A6--A8 are 266~cm in diameter and contain 240 blocks arranged 
in 5 layers, and stations A9--A11 are 306~cm in diameter and contain 240 blocks
arranged in 4 layers.
A12 is operated in air; its design is somewhat different from that of the other 
stations. 
As a result of the staggering scheme, particles incident on any station 
are intercepted by blocks in at least three rings, for a total minimum 
effective depth of 21\,$X_0$. 
Most incident particles are intercepted
by four or more blocks (27\,$X_0$).

Station A1 was constructed as a prototype during the first half of 2009.
It was installed in the NA62 beamline at CERN and tested in October 2009.
The design study and completed station are shown in \Fig{fig:station}.
Various improvements to the design were made on the basis of the results 
from the test, and station A2 was constructed and tested in the T9 beamline 
at the 
CERN PS in August 2010. A1 was then rebuilt to incorporate the design 
improvements, and construction of the remaining stations was commenced.

\section{LAV Construction}

The OPAL detector modules (lead glass block plus PMT) were manufactured 
by Hamamatsu during the mid-1980s. Their recycling requires substantial 
care throughout the assembly procedure.

After the storage area in which the modules were kept was inundated 
during a flash flood in spring 2008, the modules were subject to an 
extensive sorting and clean-up effort carried out at CERN by an industrial 
recovery firm. They were subsequently shipped to Frascati, where the LAVs
are assembled.

The glass at the interface with the stainless steel flange is 
fragile and is found to be fractured in a few percent of the modules 
upon first examination. This is attributed to thermally induced stress.
While damaged modules are discarded, the first step in the processing of 
intact modules is to reinforce the interface. Using epoxy resin, 
20~cm$^2 \times 0.5$-mm thick stainless steel plates are attached 
across the glass-steel interface on all four sides of the block. 
Calculations indicate and static tests confirm that the reinforced bond 
is several times stronger than the original bond.
After reinforcement, the blocks are unwrapped and cleaned with isopropanol 
or acetone. 
They are then wrapped with a new laser-cut and 
heat-welded Tyvek cover.

During the A1 test beam, ringing of the analog signal was observed to lead 
to errors in charge reconstruction using the time-over-threshold technique
discussed below. This problem was traced to a small parasitic inductance in 
the PMT dynodes and solved by replacing the original OPAL HV dividers
with new dividers of our own design. The new divider features 
additional resistors on the last three dynodes and anode to damp out the 
oscillation, storage capacitors for the last three dynode stages to improve 
response linearity for large signals, and a decoupling resistor between the 
HV and signal grounds to decrease noise. 
\begin{figure}
\centering
\includegraphics[height=0.165\textheight]{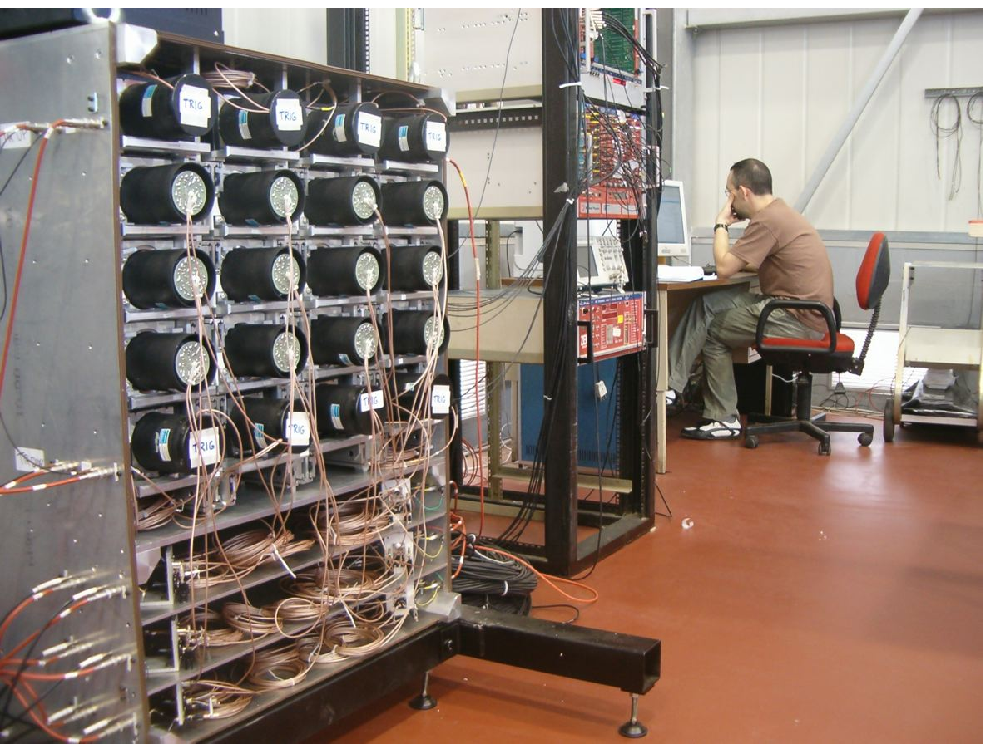}
\hspace{0.10\linewidth}
\includegraphics[height=0.165\textheight]{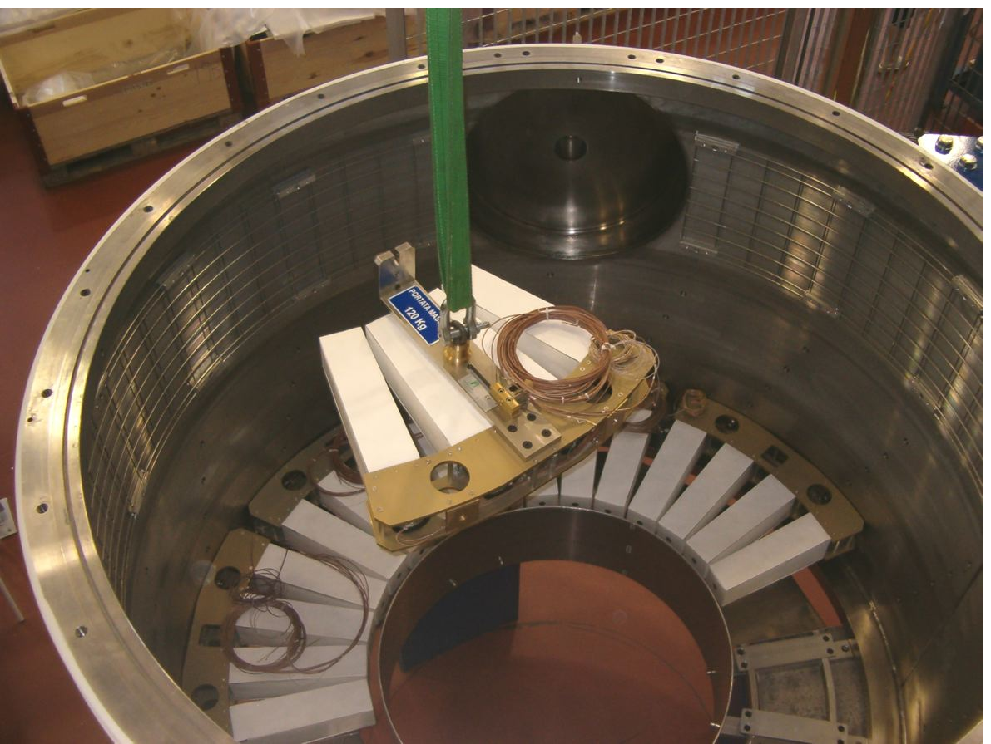}
\caption{Left: Test stand for characterization of detector 
modules. The top and bottom rows of blocks form a telescope for 
the selection of vertical cosmic rays. During operation the front
panel is closed. Right: Installation of modules into
the steel vacuum vessel.}
\label{fig:oven_assy}
\end{figure}

After the divider is replaced, the blocks are tested and characterized 
12 at a time using a test station featuring an LED pulser and cosmic-ray 
telescope (\Fig{fig:oven_assy}).
The PMT gains are measured first, by varying the
HV settings and mapping out the 
response for each block. Using the gain curves so obtained, the PMTs
are set to a reference value of the gain ($1\times10^6$),
and the response to cosmic rays selected by the telescope is measured; the
photoelectron yield for the block (p.e./MeV) is then obtained assuming that
vertically incident cosmic rays leave 77 MeV in each block. Yields of 
0.34~p.e./MeV are typical.
Finally, using the gain curves 
and the measurements of photoelectron yield, the PMT voltages are 
set to the values expected to produce a common output charge level of 4.5 pC
for cosmic-ray events. The response is measured and the HV setting is 
validated. Thus, at the end of a 12-hour cycle, which is fully automated 
using LabView, we have PMT gain and photoelectron yield measurements as well
as the operational HV settings for 12 modules. Additional data (current-draw
measurements, dark-count rates) are also collected using the test station.

The OPAL design features an optical port at the base of each 
module. Blue LEDs are installed in these optical ports
as part of the calibration and monitoring system, and will allow
monitoring of the operational status and relative timing for each 
block. A low-capacitance LED was chosen to minimize the rise and fall
times of the light pulse; this is important for use with the 
time-over-threshold-based readout system discussed in the following section.
In principle, the LED system should allow in-situ gain calibration as well. 

After testing and characterization, the blocks are arranged in groups of 
four in an aluminum mounting bracket. For the installation, the vacuum 
vessel is turned on end. The aluminum mounting bracket with four modules
is lowered into the upended vessel by overhead crane and bolted to the 
wall, as illustrated in \Fig{fig:oven_assy}, right.
  
\section{Front-End Electronics}

While the amplitude of the PMT signal from a minimum ionizing 
particle is about 20~mV, signals from 20-GeV showers may be as large as 10~V.
Our readout scheme is based on the time-over-threshold (ToT) 
technique and furnishes time and energy measurements over  
a large range of incident photon energies. The scheme is implemented 
using a dedicated front-end ToT discriminator board of our own 
design \cite{A+11:ToT} and a digital readout board (TEL62) used by 
various NA62 detector subsystems \cite{A+11:TEL62}.   
\begin{figure}
\centering
\includegraphics[height=0.225\textheight]{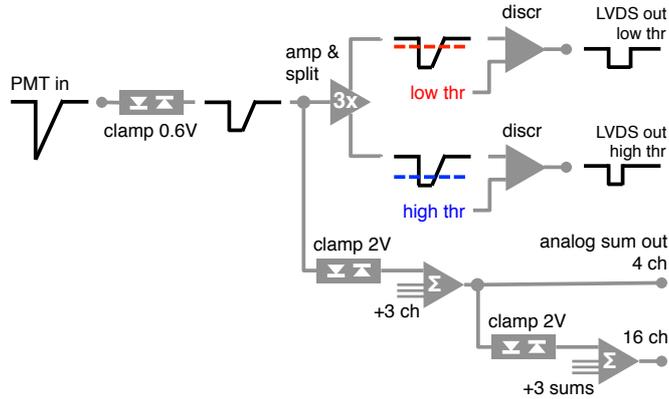}
\caption{Conceptual schematic of the ToT discriminator board.}
\label{fig:fee}
\end{figure}

The ToT discriminator converts the analog signals from 
the detector to low-voltage differential signal (LVDS) pulses, with width 
equal to the duration of the analog signal from the detector above a 
specified threshold. After a clamping stage for protection, the signal
from each PMT is passively split.
One copy of the clamped analog signal is summed with the signals 
from adjacent channels and made available via a front-panel connector
for diagnostic purposes.
The other copy is amplified $\times3$ and used as input to a 
two different comparators, corresponding to a low threshold
and a high threshold. Each comparator generates a separate LVDS output.
We expect to use thresholds of about 5 and 50~mV against a noise level of 
about 2 mV; the threshold values are programmable.
A conceptual schematic of the ToT discriminator board is presented in 
\Fig{fig:fee}.
The production model has 32 input channels and 64 output channels.

The TEL62 is based on the design of the TELL1 readout board developed for the 
LHCb experiment. On the TEL62, TDC mezzanines measure the leading and 
trailing times of the LVDS pulses. Using a time-to-charge calibration 
parameterization, the FPGA on board the TEL62 calculates the time 
corrected for slewing, as well the charge for each hit as reconstructed 
from the pulse width above threshold. This information is sent to the 
subsequent DAQ stages. Level-0 trigger primitives are also calculated on 
board the TEL62 and sent to the level-0 trigger processor.

\section{Test-Beam Performance}

LAV A2 was tested using the positive secondary beam in the T9
area at the CERN PS in August 2010. Data were collected at various beam
momenta over the interval 0.3--10~GeV. 
The composition of the T9 beam changes as a function of the momentum setting.
At 0.3~GeV, the beam is roughly 70\% $e^+$ and 30\% $\pi^+$ (including 
decay $\mu^+$), while above
6~GeV the $e^+$ component drops off sharply.
Two threshold CO$_2$ Cerenkov counters in the beamline allowed samples 
enriched $e^+$ and $\mu^+$ to be selected. The beam focus coincided 
approximately with A2's first layer of lead glass blocks. 
Two crossed scintillator paddles were placed at the beam entrance, 
and one larger paddle was placed at the beam exit. 
A prototype of the ToT discriminator board was used for the test, 
mainly in conjunction with commercial QDCs and TDCs.

\begin{figure}
\centering
\includegraphics[height=0.21\textheight]{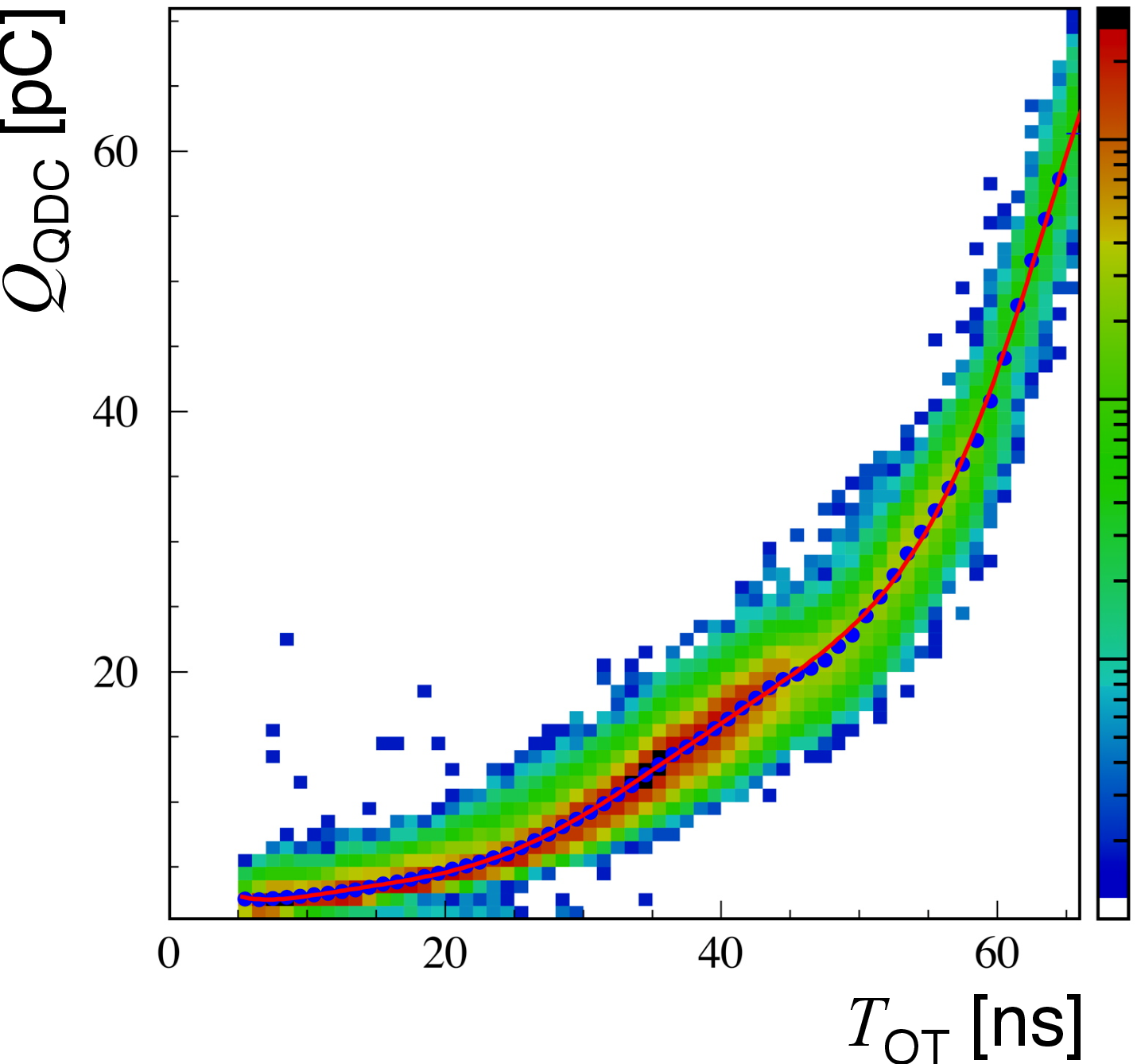}
\hspace{0.10\linewidth}
\includegraphics[height=0.21\textheight]{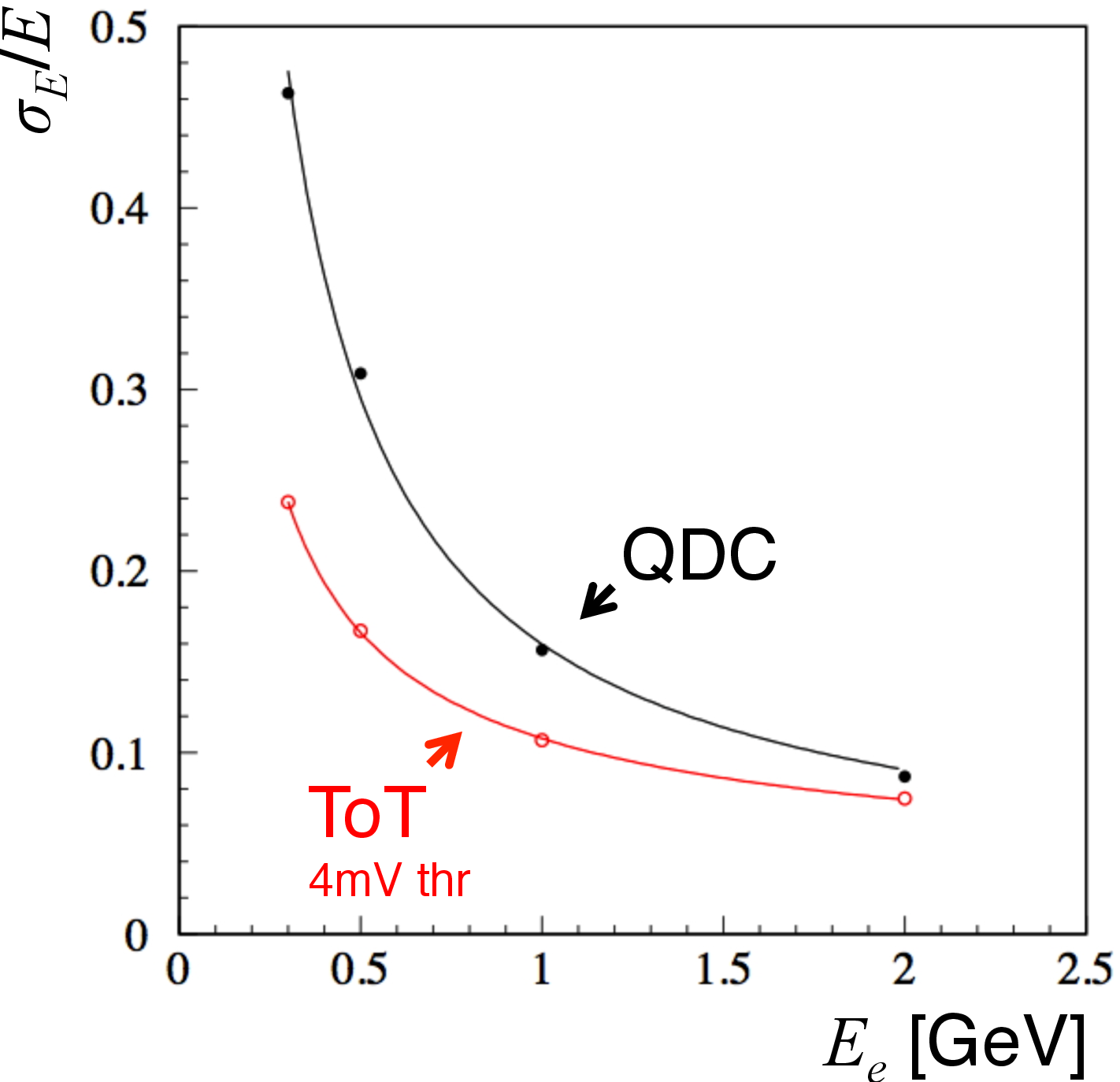}
\caption{Left: Scatter plot of signal charge measured using QDCs vs.\ 
signal ToT, for electrons of various energies. 
Right: Energy resolution obtained using ToT 
compared with that obtained using QDCs, as a function of incident 
electron energy.}
\label{fig:qtot_eres}
\end{figure}
Figure~\ref{fig:qtot_eres} shows a scatter plot of the signal charge 
measured using the QDCs vs.\ the signal time over a threshold of 4~mV, for 
electrons. The data are summed for all blocks on which the beam was 
incident and over runs of different energies. For small signals, a
small increase in the integrated charge corresponds to a large increase 
in the time over threshold; the ToT measurement 
provides more sensitivity for small signals than does the QDC measurement.
This is a desirable property for the LAV detectors, since high detection 
efficiency is required for low-energy photons.
Parameterizations of the type illustrated in \Fig{fig:qtot_eres}
are used to convert ToT to an effective charge measurement.
The energy resolution obtained using the ToT technique is compared with that 
obtained using the QDCs in \Fig{fig:qtot_eres}. As expected from the form of 
the curve in \Fig{fig:qtot_eres}, at low energies,
the resolution obtained with the ToT technique is better than that obtained 
with the QDCs. The fits in \Fig{fig:qtot_eres} give
\begin{eqnarray*}
{\rm QDC:}&\hspace{1mm}& \sigma_E/E = 8.6\%/\sqrt{E~{\rm[GeV]}} \oplus 13\%/E, \\
{\rm ToT:}&\hspace{1mm}& \sigma_E/E = 9.2\%/\sqrt{E~{\rm[GeV]}} \oplus 5\%/E \oplus 2.5\%.
\end{eqnarray*}
While, as expected, the statistical contribution to the energy resolution 
is about the same with either readout scheme, the contribution from noise 
(term proportional to $1/E$) appears to be significantly smaller with the ToT 
technique. The presence of the constant term with the ToT technique may be 
due to small differences in the charge vs.\ ToT curves from block to block.

The width of the signal time distribution in slices of charge gives a 
measurement of the time resolution. A fit to the measurements of $\sigma_t$ 
vs. charge from ToT (4~mV threshold) for a single 
block gives $\sigma_t = 220~{\rm ps}/\sqrt{E~{\rm[GeV]}} \oplus 140~{\rm ps}$, 
where the constant term is assumed to be due to trigger jitter.
The intrinsic time resolution of the detector can be estimated from 
the width of the distribution of signal time differences for two 
successive blocks.
Assuming that the two blocks have the same intrinsic time response and that 
there is no common-mode contribution to the resolution,
$\sigma_{t_1} = p/\sqrt{E_1}$ and $\sigma_{t_2} = p/\sqrt{E_2}$,
so that we expect $\sigma_{\Delta t} = p/[E_1 E_2/(E_1 + E_2)]^{1/2}$, with no
constant term. Our measurements are consistent with $p = 210$~ps
and no constant term.

\begin{figure}
\centering
\includegraphics[height=0.19\textheight]{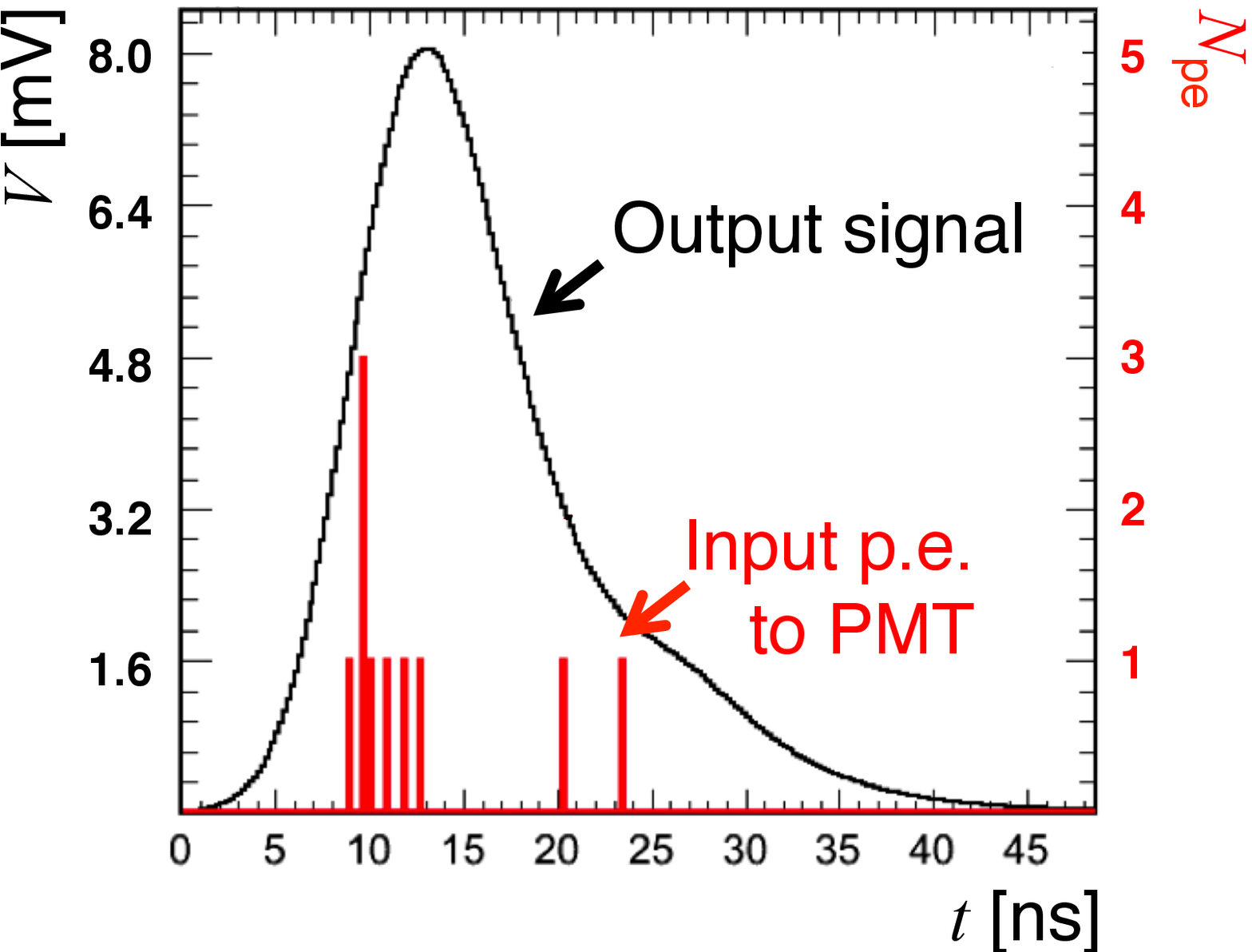}
\hspace{0.05\linewidth}
\includegraphics[height=0.19\textheight]{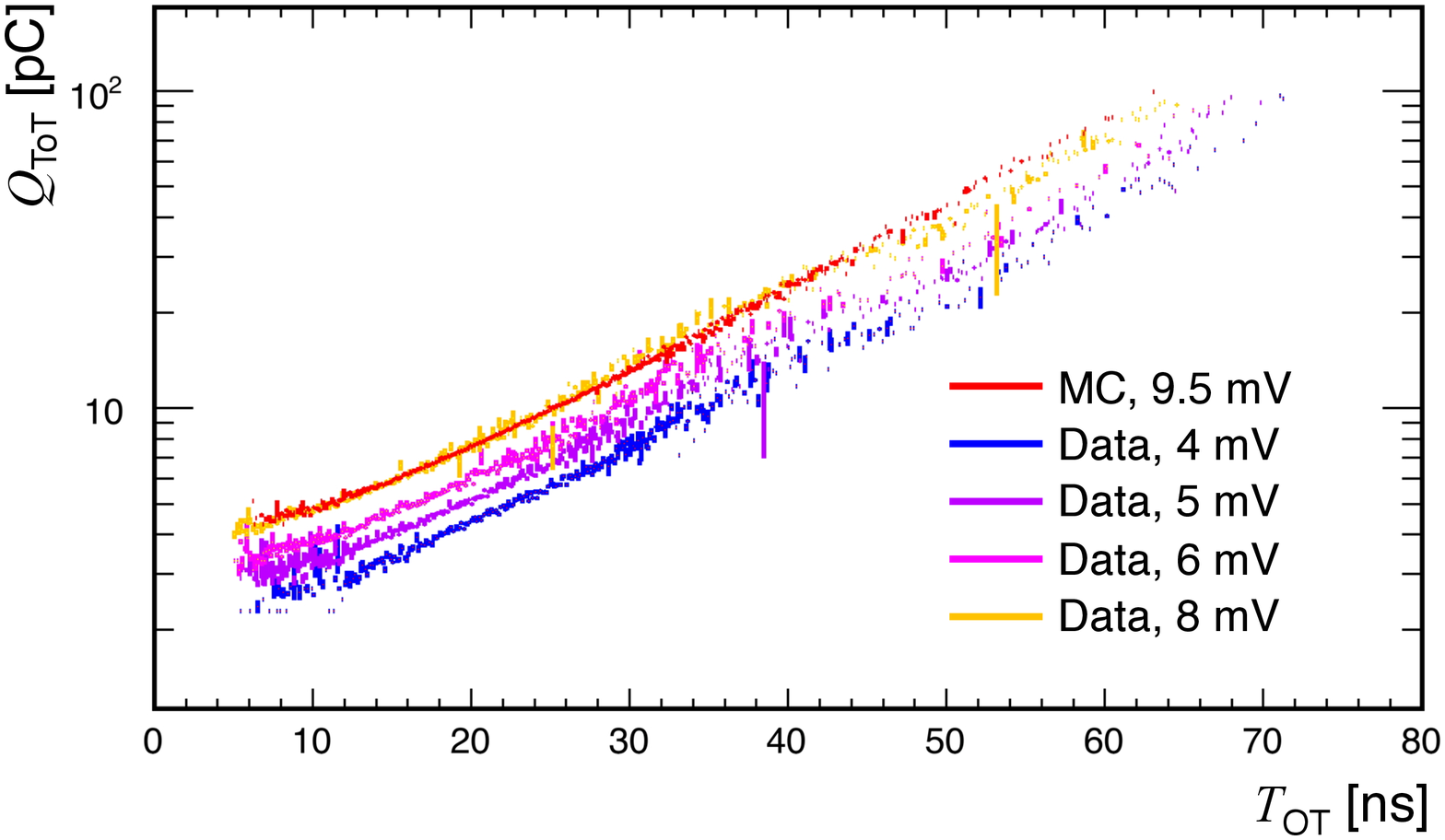}
\caption{Left: Simulation of PMT signals. Number of 
photoelectrons produced in bins of time (red, scale at right) and 
resulting PMT signal (black, scale at left). Right: Comparison of 
charge vs.\ ToT curves for MC (9.5 mV threshold) and data (various 
thresholds).}%
\label{fig:sig_sim}
\end{figure}
The Geant4-based NA62 Monte Carlo (MC) simulation includes a detailed 
description of the LAV geometry and materials. The arrival times of the 
Cerenkov photons at the photocathode are simulated as shown by the 
histogram in red in \Fig{fig:sig_sim}. A complete simulation of
the PMT uses this information to generate an output signal, taking into 
account the PMT gain and transit-time fluctuations, the capacitance of the 
PMT, and dispersion in the readout cable. The resulting signal, which
is illustrated by the black curve in \Fig{fig:sig_sim}, is input to 
a full simulation of the ToT discriminator.

Figure~\ref{fig:sig_sim} shows charge vs.\ ToT curves for test-beam muons for
four different threshold settings (4, 5, 6, and 8~mV). The red curve is 
from the simulation, with a threshold at 9.5~mV and all other adjustable 
parameters set to typical or measured values. The simulation with a 
threshold of 9.5~mV accurately reproduces the shape of the measured charge 
vs.\ ToT curve with threshold 8~mV. This slight discrepancy is easily 
attributed to the accuracy of the manual threshold adjustment on the 
prototype ToT discriminator board. These results thus confirm our detailed 
understanding of the detector and the readout chain. 

\section{Status and Outlook}

As of July 2012, LAVs A1 through A8 have been completed and installed in 
the beamline. An additional station, A11, is nearly complete at Frascati.
Serial production of the front-end electronics boards is underway. 
As the front-end electronics become available during summer 2012, 
LAVs A1 though A8 will be read out without beam. They will be fully 
tested in an NA62 technical run scheduled for November 2012.
NA62 data taking with all detectors installed is planned for early 2014.

\section*{References}

\bibliographystyle{iopart-num}
\bibliography{calor_proc}

\end{document}